\documentclass[fleqn,usenatbib]{mnras}

\usepackage{newtxtext,newtxmath}

\usepackage[T1]{fontenc}
\usepackage{ae,aecompl}

\usepackage{graphicx}	
\usepackage{amsmath}	
\usepackage{amssymb}	
\usepackage{epsfig}
\usepackage{epstopdf}

\def\be{\begin{equation}}
\def\ee{\end{equation}}

\title[Mass assembly history and $H_0$ tension]{Mass assembly history of dark matter halos in the light of $H_0$ tension}

\author[Hamed Kameli, Shant Baghram]{
Hamed Kameli 
~and Shant Baghram \thanks{baghram@sharif.edu}
\\
Department of Physics, Sharif University of Technology, P.~O.~Box 11155-9161, Tehran, Iran\\\
}

\date{Accepted XXX. Received YYY; in original form ZZZ}

\pubyear{2020}

\begin{document}
\label{firstpage}
\pagerange{\pageref{firstpage}--\pageref{lastpage}}
\maketitle

\begin{abstract}
The Hubble tension may introduce a new course of action to revise the standard $\Lambda$CDM model to unravel dark energy and dark matter physics.
The Hubble parameter can be reconstructed by late-time observations of the background evolution model independently.
We relate the reconstructed Hubble parameter to the structure formation and large scale structure observables in this work. We use the excursion set theory to calculate the number density of dark matter halos and the mass function of progenitors. We obtain the results for both the Markov and non-Markov extension of the excursion set theory  in the context of spherical and ellipsoidal collapse.
We show that the number density of dark matter halos in the reconstructed model has approximately  $\sim2\sigma$ difference in comparison to the Planck-2018 $\Lambda$CDM in the mass range of $M\gtrsim10^{12}M_{\odot}$. We also compare the dark matter halo progenitor mass function with the pair-galaxy statistics and their mass assembly history from observational data of the HST, CANDEL survey. Due to complications to distinguish the ratio of accretion and merger in mass assembly, our result on pair fraction is for illustration only. However, a $\sim5$ times more accurate observations will be promising to distinguish the reconstructed model and the Planck-2018 $\Lambda$CDM.
\end{abstract}
\begin{keywords}
cosmology: large-scale structure of Universe,  cosmology: dark matter,   galaxies: haloes
\end{keywords}


\section{Introduction}
\label{Sec:1}
It is almost more than two decades since the discovery of the accelerated expansion of the Universe with the observation of supernova type Ia (SNe Ia) \cite{Riess:1998cb, Perlmutter:1998np}. The standard model of cosmology known as $\Lambda$CDM emerged and withstood with most recent observations. The precise measurement of the statistics of the cosmic microwave background (CMB) radiation fluctuations enables us well to constrain the standard model parameters \cite{Aghanim:2018eyx}. On the other hand, late-time observations of large-scale structure (LSS), such as statistics of galaxy clustering \cite{Percival:2009xn, Alam:2016hwk, Camacho:2018mel} and weak lensing \cite{Hildebrandt:2016iqg} are prominent examples, which are in good agreement with the standard model. Despite all these successes, the nature of dark energy (DE), dark matter (DM), and the physics of the early Universe is still unknown \cite{Bull:2015stt}. One possible way to address these fundamental questions is to focus on the data and theory's known tensions through new ideas \cite{Peebles:2012mz}. One of the main tensions comes from the measurement of the Hubble constant $H_0$. The $H_0$ obtained from local standard candles has almost $\sim4.4 \sigma$ difference with the CMB data \cite{Riess:2019cxk} \footnote{$H_0=74.03\pm1.42 ~ \text{km s}^{-1}\text{Mpc}^{-1}$ from local measurement \cite{Riess:2019cxk} and $H_0=67.27\pm 0.60 ~\text{km s}^{-1}\text{Mpc}^{-1}$ from Planck-2018 data \cite{Aghanim:2018eyx}}. This discrepancy could result from a statistical fluke, observational systematics, or a hint to a new physics. In this direction, many proposals have been introduced, such as early time modification of sound horizon \cite{Poulin:2018cxd}, late time DE models \cite{DiValentino:2020naf}, interacting DE-DM models \cite{DiValentino:2019ffd}, and modified gravity theories \cite{Khosravi:2017hfi}. For a new and comprehensive review see \cite{DiValentino:2021izs}.
Obviously, a new physics proposed to solve the Hubble tension should also be consistent with other cosmological observations. The LSS observations are important to study beyond standard $\Lambda$CDM models \cite{Ishak:2005zs,Pogosian:2007sw, Bertschinger:2008zb, Baghram:2009fr,Baghram:2010mc,Baghram:2014qja,Koyama:2015vza,Klypin:2020tud}.\\
In this work, we suggest testing the effect of the background evolution of the late time Universe, encoded in the Hubble parameter, on the LSS observations. We study the effect of this modified Hubble parameter on the matter power spectrum, number density, and the progenitor history of DM halos in the excursion set theory context \cite{Lacey:1993iv,Sheth:1999mn,Sheth:2001dp}. We show that the modified Hubble parameter has observable effects on DM halos' statistics in the context of hierarchical structure formation. The modification in the progenitor history of DM halos can be considered as a new proposal to address the questions and caveats in structure formation such as the seeds of supermassive black holes \cite{Heckman:2014kza}, the quenching of massive galaxies \cite{man2018star} and the observation of flat galaxies \cite{Peebles:2020bph}. \\
The structure of this work is as below: In Sec.\ref{Sec:2}, we review the theoretical background of this work, specially the excursion set theory (EST) and the LSS observations in linear and non-linear scales. In Sec.\ref{Sec:3}, we discuss our results and implications of reconstructed Hubble parameter (obtained from \cite{Wang:2018fng}) in LSS observables and finally in Sec.\ref{Sec:4} we conclude and propose the future remarks. The results for flat $\Lambda$CDM model are based on Planck-2018 with matter density of $\Omega_m=0.27$, the $H_0 =67\, \text{km s}^{-1}\text{Mpc}^{-1}$ and $n_s = 0.96$ \cite{Aghanim:2018eyx}.
\section{ Theoretical background: From Hubble parameter to LSS observables}
\label{Sec:2}
This section reviews the theoretical background of this work, which shows the effect of the Hubble parameter on DM halos formation history and LSS observables such as halos number density and progenitor history. First, we discuss the linear theory in the standard model. Then the non-linear structure formation of DM halos is discussed in the context of EST for spherical collapse (SC) and ellipsoidal collapse (EC). Finally, we review the recent implications of the non-Markov EST model and the numerical counting methods to obtain the halos number density and their mass assembly history.
\subsection*{Linear theory}
To study the linear structure formation, we use the perturbed Friedmann Lemaitre Robertson Walker (FLRW) metric
\be
ds^2=a^2(\eta)\left[-(1+2\Psi(t,\vec{x}))d\eta ^2 + (1+2\Phi(t,\vec{x}))dx^idx^j\delta_{ij}\right],
\ee
where $\eta$ is the conformal time, $\Psi$ is the Newtonian potential and $\Phi$ is the curvature perturbation. Using the Einstein's equations we have relativistic Poisson equation \cite{Amendola:2003wa}
\be \label{eq:Poisson}
k^2\Phi(k,z)=4\pi G (1+z)^{-2}\bar{\rho}(z)[\delta(k,z) + \frac{3{\cal{H}}(1+w)\theta(k,z)}{k^2}],
\ee
where $\bar{\rho}$ is the mean matter density, $\delta=\rho/\bar{\rho} - 1$ is density contrast and $\theta = ikv_k$ is the peculiar velocity's divergence in Fourier space. Note that $\cal{H}$ is the conformal Hubble parameter and $w=P/\rho$ is pressure to density ratio (we set w=0 as we deal with non-relativistic DM). The continuity and Euler equations for cold DM raised from energy-momentum conservation are
\be \label{eq:mass}
\delta' = - \theta - 3\Phi',
\ee
\be \label{eq:Euler}
\theta' + {\cal{H}}\theta = k^2\Psi.
\ee
where $\prime$ is derivative with respect to the conformal time. Combing equations(\ref{eq:Poisson},\ref{eq:mass},\ref{eq:Euler}), we find the DM density contrast evolution in terms of redshift in sub-horizon scales ($k\gg {\cal{H}}$) and in quasi-static regime (ignoring the time derivatives of Bardeen potentials in comparison with Hubble time scale) as
\be \label{eq:growth}
\frac{d^2\delta}{dz^2} + [\frac{dE(z)/dz}{E(z)}-\frac{1}{1+z}]\frac{d\delta}{dz} - \frac{3}{2}\Omega_m\frac{1+z}{E^2(z)}\delta = 0,
\ee
where $E(z)=H(z)/H_0$ is normalized Hubble parameter, $\Omega_m$ is the matter density parameter.
{{The growth function $D(z)$ is defined as
\be \label{eq:deltac}
\delta(z) = {D(z)}\delta_{0},
\ee 
\begin{figure} 
	\includegraphics[scale=0.5]{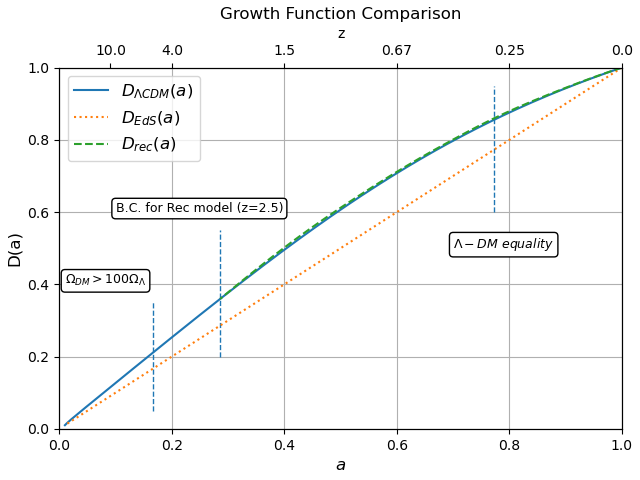} 
	\caption{The growth function is plotted versus scale factor (redshift in upper x-axis) for $\Lambda$CDM (solid blue line), EdS (red dotted line) and reconstructed model (green dashed line) by solving  \ref{eq:growth} for $D(z)$ with appropriate boundary condition discussed in the main text. } \label{fig0}
\end{figure} 
which is normalized to unity at redshift $z=0$. In Fig. \ref{fig0}, we show the solution of the  equation \ref{eq:growth} for $\Lambda$CDM and reconstructed model (introduced in upcoming section). The  differential equation for growth function is solved with boundary conditions as the growth function scales with scale factor ($D(a)=a$) in deep dark matter era and it is normalized to unity in present time. The results are compared with Einstein–de Sitter (EdS) solution (for similar result see figure 1 in \cite{Huterer:2013xky}).}}
The growth function incorporates only the time evolution of the density contrast. Accordingly, the linear matter power spectrum $P_L(k,z)$ will be defined as
\be
P_L(k,z) = A_l k^{n_s}D^2(z)T^2(k),
\ee
where $A_l$ is the late time amplitude of perturbations, $n_s$ is the spectral index of the perturbations, and $T(k)$ is the transfer function. The transfer function introduces the scale-dependence of Bardeen potentials' evolution, considering the physics of the equality era and horizon entry. We use the Eisenstein-Hu Transfer function \cite{Eisenstein:1997ik}. {{Note that the amplitude of the power spectrum is fixed by $\sigma_8$, which is the variance of the perturbations in $R=8$ Mpc/h in $z=0$.}}
\subsection*{Non-linear structure formation}
One of the main observables in non-linear structure formation is the luminosity distribution of galaxies, which are tightly related to the number density of DM halos \cite{Cooray:2002dia}. An old but sophisticated way to calculate the number density of DM halos is the idea of Press-Schechter (PS) \cite{Press:1973iz}. PS formalism proposed that probability distribution function (PDF) of the density contrast in high redshifts, where the perturbations are almost Gaussian and linear, can be used to predict the late time number density of the DM halos. The PDF fraction of linear density contrast with a larger value than the spherical collapse barrier \cite{Gunn1972} is considered the fraction of the gravitationally bound objects with the same amount of mass enclosed in the initial smoothing radius \cite{Zentner:2006vw}. Later on, \cite{Bond:1990iw} introduced the Excursion Set Theory (EST), which relates the statistical properties of the initial density contrast field to the number density of structures by using the stochastic process techniques. 
A set of trajectories is plotted in the 2-dimensional plane of density contrast versus variance. The trajectory steps are generated by smoothing window function around arbitrary points in initial density contrast field \cite{Cooray:2002dia,Zentner:2006vw,Nikakhtar:2016bju}. The statistics of the first up-crossing from a specific barrier
 of these random walk trajectories $f_{\text{FU}}$ is related to the number density of DM halos $n(M)$ as
\be
n(M)dM= \frac{\bar{\rho}}{M}f_{\text{FU}}(S)|\frac{dS}{dM}|dM,
\ee
where $f_{\text{FU}}$ is the first up-crossing counts of density contrast variance in the interval of $S$ and $S+dS$. The variance in each smoothing scale $R$ is obtained from the weighted integral of linear matter power spectrum as
\be \label{eq:variance}
S(R)=\sigma^2(R)=\frac{1}{2\pi^2}\int dk k^2P_L(k,z=0)\tilde{W}^2(kR),
\ee
{{where $\tilde{W}(kR)$ is the Fourier transform of window (smoothing) function in real space. In the EST framework, the variance is independent of redshift, so the mass is related to the variance in $z=0$ with $D(z=0)=1$.}}\footnote{ We use the Gaussian filter $\tilde{W}(kR)=\exp[{-(kR)^2/2}]$ for non-Markov EST.} \cite{Bond:1990iw} show for sharp k-space window function, the trajectories execute a Markov random walk. Accordingly, for Markov trajectory, the analytical expression for the first up-crossing distribution is
\be
f_{\text{FU}}(S,\delta_c(z))dS=\frac{1}{\sqrt{2\pi}}\frac{\delta_c(z)}{S^{3/2}}e^{-\frac{\delta^2_c(z)}{2S}}dS, \label{eq:ffu}
\ee 
{{where the redshift dependence appeared in $\delta_c(z)=\delta_c  / D(z)$ via the growth function. Note that we use the constant critical density (barrier in EST) $\delta_c=1.686$ for spherical collapse.}} \footnote{ The cosmology dependence of the barrier for spherical collapse is discussed in \cite{Mo2010,Courtin:2010gx}. However, the effect is negligible and in this work, we use the constant value of the  Einstein-de Sitter (EdS) universe for spherical collapse.}
In more realistic models the barrier could be both scale dependent (such as ellipsoidal collapse) and stochastic to describes the effect of shear and velocity dispersion \cite{Robertson:2008jr}. In this direction, we use the moving barrier for density contrast as \cite{Sheth:2001,Sheth:2001dp} to study the EC.
\be \label{eq:moving}
\delta_{\text{ec}}(S,z) = \sqrt{a}\delta_c(z)[1+\beta (a\nu^2)^{-\alpha}],
\ee
\begin{figure} 
	\includegraphics[scale=0.5]{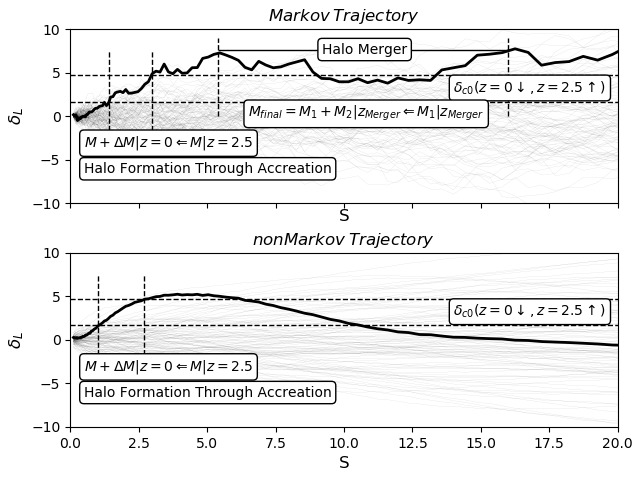} 
	\caption{ Top panel: A set of Markov trajectories. Bottom panel: a set of non-Markov trajectories. The idea of the halo formation and merger is depicted in the figure.} \label{fig1}
\end{figure} 
{{where $\delta_{\text{ec}}(S,z)$ is the moving barrier of EC which depends on variance and redshift and $\nu(z)=\delta_c/(\sigma(z=0) D(z))$}}. Note that we use the constants $a=0.7$, $\alpha=0.615 $ and $\beta=0.485$.\\
It worths mentioning using other window functions leads to non-Markov trajectories. Note that there is no analytical expression for $f_{\text{FU}}$ in non-Markov case. So the first up-crossing of the trajectories should be counted numerically (for more discussion see \cite{Nikakhtar:2018qqg}). \\
In this context the mass assembly of DM halos can be studied by conditional up-crossing. The  conditional mass distribution at redshift $z_2$ of DM halos having mass $M_1$ at redshift $z_1$ in the context of spherical collapse and Markov EST (corresponding to variance $S_1$ via equation(\ref{eq:variance})) at redshift $z_1$ to a DM halo with a larger mass $M_2>M_1$ (corresponding to $S_2$) at redshift $z_2<z_1$ is
\be \label{eq:ffuc}
f_{\text{FU}}(S_1,\delta_1|S_2,\delta_2)=\frac{1}{\sqrt{2\pi}}.\frac{\delta_c(z_1) - \delta_c (z_2)}{(S_1-S_2)^{3/2}} e^{-\frac{(\delta_c(z_1) - \delta_c(z_2))^2 }{2(S_1-S_2)}}.
\ee 
The above conditional mass function can be obtained by numerical counting method (described in next subsection) for elliptical collapse model and also non-Markov extension of EST.\\
In Fig.\ref{fig1}, we plot a pedagogical figure with a constant barrier (SC) for a set of Markov trajectories in the top panel and a non-Markov set (see next subsection) in the bottom panel. We show the idea of halo formation through accretion (a smooth increase of density contrast in terms of variance) in the redshift interval of $z=0-2.5$, and halo merger (a visible jump in trajectories in specified redshift for two different masses) in the upper panel. {{ Equation (\ref{eq:ffuc})  shows the contribution of the both merger and accretion to progenitor mass assembly.}} Note, that in the EC case, the trajectories are the same, but the barrier is a moving one defined by equation(\ref{eq:moving}). In the EC, the mergers are defined by jumps of trajectories which cross the same $\delta_{\text{ec}}(S,z)$ curve with the specific redshift $z=z_*$ \cite{Moreno:2007wu}.
\subsection*{Non-Markov extension of EST}
\begin{figure}
\includegraphics[scale=0.5]{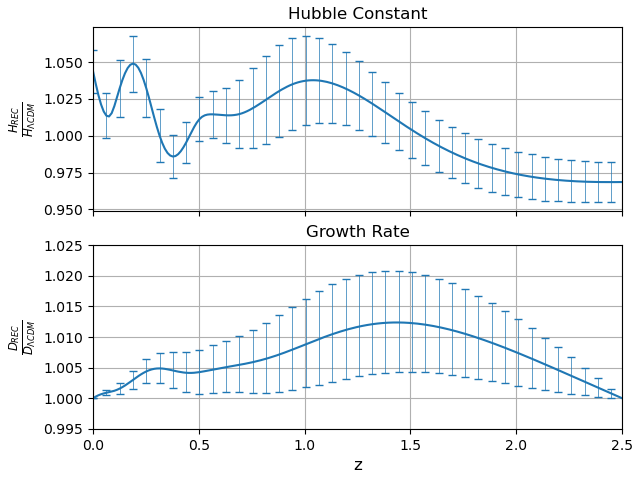}
\caption{Top panel: the reconstructed Hubble parameter normalized to Planck-2018 $\Lambda$CDM is plotted versus redshift. The data used in the reconstructed model are SNe Type Ia, BAO and Planck distance indicator (see the main text for the references). The error-bars are the $1\sigma$ confidence level of the reconstructed Hubble parameter. Bottom panel: the ratio of the growth rate of the reconstructed model to $\Lambda$CDM prediction with  the $1\sigma$ confidence level is plotted.} \label{fig2}
\end{figure}
To study DM halos' number density and their mass assembly history, we need more realistic models and better approximations. One of the main caveats of standard EST is the use of the k-space sharp window function. If we choose more realistic smoothing functions, such as real space top-hat or Gaussian window functions (as used in this work), we end up with non-Markov trajectories. There are many attempts to address the problem of the first up-crossing in non-Markov walks. (For-example see \cite{Maggiore:2009rv,Maggiore:2009rw,Musso:2012qk,Musso:2013pha} for accurate approximation).\\
We use the numerical method developed in \cite{Nikakhtar:2018qqg, Baghram:2019jlu} and its extension to calculate number density and mass assembly probability in $\Lambda$CDM cosmology \cite{Kameli:2019bki}.
In non-Markov trajectories, the height of the smoothed density field extrapolated to the present time $\delta_R(\vec{x})$ is correlated to the density contrast in previous variance steps. These correlations lead to a smoother trajectories in comparison to jagged ones in Markov case (see Fig.\ref{fig1}), which changes the statistics of DM halos. This means that the density contrast in the $n-$th step can be written as
\begin{equation}
\delta_n=\langle \delta_n|\delta_{n-1},...,\delta_1\rangle +\sigma_{n|n-1,...,1}\xi_n,
\end{equation}
where the first term indicate that the height in $n-$th step of variance depends on the previous steps and $\xi_n$ is a zero mean, unit variance Gaussian random number ($\langle \xi_n\xi_m \rangle = \delta_{mn}$). In \cite{Nikakhtar:2018qqg}, a numerical method based on Cholesky decomposition is introduced to generate an ensemble of trajectories with the correct statistical  characteristic encoded in correlation matrix $C_{ij}$ as
\be \label{eq:cij}
\langle \delta_i \delta_j \rangle \equiv C_{ij}=\int \frac{dk}{k} \frac{k^3P_L(k)}{2\pi^2} \tilde{W}(kR_i)\tilde{W}(kR_j),
\ee 
where $i(j)$ is related to the smoothing scale $R_i(R_j)$, so the statistical correlation of density contrast in different scales is embedded in $C_{ij}$.Then the non-Markov trajectories are obtained from
\be
\delta_i=\sum_{j}L_{ij}\xi_j,
\ee
where $L_{ij}$ are the components of lower triangular matrix related to the decomposed $C={\bf{L}}{\bf{L}}^T$. Note that $\xi_{j}$, is a random number with Gaussian distribution. By using a proper  power spectrum in equation(\ref{eq:cij}), we can use the Cholesky decomposition method to produce cosmological model dependent trajectories. Note that the first up-crossing problem can be estimated numerically in the non-Markov case with both constant barrier for SC and moving barrier for EC. In this works, for the first time, we show the number density (conditional number density) results of non-Markov with moving barrier case. Accordingly, our proposed results in the next section considered both effects of a more realistic smoothing window function and collapse model.
In the next section, we will use the Cholesky method to produce the trajectories with the standard  Planck-2018 $\Lambda$CDM and the reconstructed Hubble parameter model.
\section{Results: From reconstructed Hubble parameter to mass assembly history }
\label{Sec:3}
\begin{figure}
\includegraphics[scale=0.5]{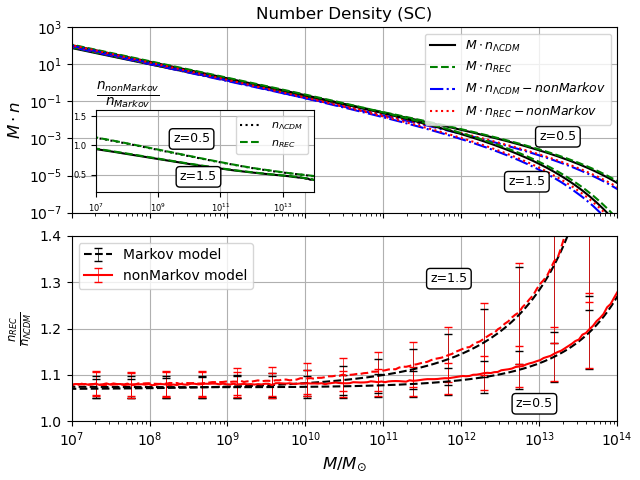}
\caption{ Top panel: The number density of dark matter halos is plotted for standard Planck-2018 $\Lambda$CDM and reconstructed models. Inset figure: We show the ratio of number density for non-Markov to Markov in both cosmological models. Bottom panel: The ratio of the number density of the reconstructed to $\Lambda$CDM models is plotted versus mass for Markov to Markov and non-Markov to non-Markov cases, separately. All the figures are plotted for spherical collapse model  in  $z=0.5$ and $z=1.5$.} \label{fig3}
\end{figure}
\begin{figure}
\includegraphics[scale=0.5]{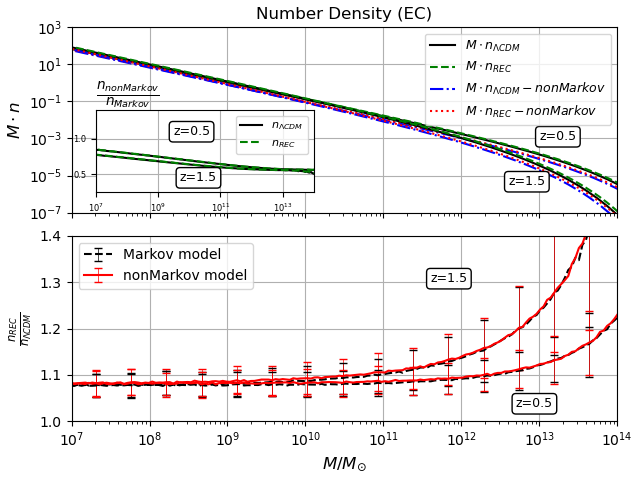}
\caption{  Top panel: The number density of dark matter halos is plotted for standard Planck-2018 $\Lambda$CDM and reconstructed models. Inset figure: We show the ratio of number density for non-Markov to Markov in both cosmological models. Bottom panel: The ratio of the number density of the reconstructed to $\Lambda$CDM models is plotted versus mass for Markov to Markov and non-Markov to non-Markov cases, separately. All the figures are plotted for ellipsoidal collapse model  in  $z=0.5$ and $z=1.5$. } \label{fig3ec}
\end{figure}
In this section, we present our results on the effect of the reconstructed Hubble parameters on the LSS observables, and we compare them with standard Planck-2018 $\Lambda$CDM predictions.
We use the late time distance indicator observations to reconstruct the Hubble parameter, independent of any proposed cosmological model introduced in \cite{Wang:2018fng}. The background data used for Hubble reconstruction are supernovae data from joint light analysis (JLA) sample \cite{Betoule:2014frx}, baryonic acoustic oscillation (BAO) measurements from 6dF Galaxy Survey (6dFGS) \cite{beutler33666df}, SDSS DR7 Main Galaxy Sample (MGS) \cite{ross2015clustering}, tomographic BOSS DR12 (TomoBAO) \cite{Wang:2016wjr}, eBOSS DR14 quasar sample (DR14Q) \cite{Ata:2017dya} and the Lyman-$\alpha$ forest of BOSS DR11 quasars \cite{Font-Ribera:2013wce}; \cite{Delubac:2014aqe}.
In Fig.\ref{fig2} top panel, the reconstructed Hubble parameter normalized to the Planck-2018 $\Lambda$CDM is plotted\cite{Wang:2018fng}.
The modified growth function is extracted by equation(\ref{eq:growth}) using the reconstructed Hubble parameter. 
{{The boundary condition of the growth function in reconstructed model is set equal to $\Lambda$CDM value at $z=2.5$ and  equal to unity in present time. 
 The reconstructed growth function ratio to the Planck-2018 $\Lambda$CDM is plotted in Fig.\ref{fig2} bottom panel. It is worth mentioning that in the context of the non-linear structure formation and EST, modified growth function can affect the matter distribution. This modification results from the variation of the redshift dependence of the barrier $\delta_c(z)$. Accordingly, the first up-crossing and conditional one (equations(\ref{eq:ffu}) and (\ref{eq:ffuc})) will be changed due to modified barrier $\delta_c$ ($\delta_{\text{ec}}(z,S)$ for EC). In this direction, we study the statistics of DM halos in both cosmological models.}}  \\
 {{The ratio of the number density of dark matter halos in two cosmological models in $z=0$ is equal to ratio of background matter density ($n_{Rec}/ n_{\Lambda CDM} = \bar{\rho}_{Rec} / \bar{\rho}_{\Lambda CDM}$). This is because the growth function is normalized to unity in both models in present time. However,  the ratio in higher redshift become more complicated due to redshift dependence of the critical density through the growth function $\delta_c(z) = \delta_c / D(z)$. }}
  {{ In Fig.\ref{fig3}, we show the number density of DM halos at $z=0.5$ and $z=1.5$ of the reconstructed and $\Lambda$CDM models for both Markov and non-Markov EST extension for spherical collapse (SC). The bottom panel shows the number density ratio in reconstructed Markov (non-Markov) to the Plank-2018 $\Lambda$CDM Markov (non-Markov) model, separately. It worth mentioning that $\delta_c(z=0)=1.686$ for both reconstructed and $\Lambda$CDM models as we assume that the process of spherical collapse is the same for both models. The differences in $D(z)$ are shown in Fig. \ref{fig0}, where the growth function is normalized to unity in present time $D(z=0)=1$. This means the number density in two models in $z=0$ is different, only through the effect of the background matter density. However, in higher redshift we can see the effect of the growth function difference in both models.}}
  In this direction, the reconstructed model can be approximated by dark energy models with a redshift dependent equation of state $w=w(z)$. Accordingly, It is shown that the critical density in dark energy models depends weakly on cosmological parameters \cite{Pace2010}. \\
In Fig.\ref{fig3ec}, we plot the same quantities as in Fig.\ref{fig3} for ellipsoidal collapse (EC), with moving barrier introduced in equation(\ref{eq:moving}). The error bars introduced due to the reconstruction method are small enough to distinguish the two models by their DM halo number density prediction in almost $\sim 2\sigma$ in DM halo mass ranges $M\gtrsim10^{12}M_{\odot}$. The Figs. \ref{fig3},\ref{fig3ec} show that the behaviour of the ratio of reconstructed to the standard model is almost the same in Markov and non-Markov cases. The ratio of the reconstructed to the standard model is almost the same for SC and EC in both Markov and non-Markov cases. This is because of the cancellation effect in the numerator and denominator of all models (both EST and collapse models).\\
\begin{figure}
	\includegraphics[scale=0.5]{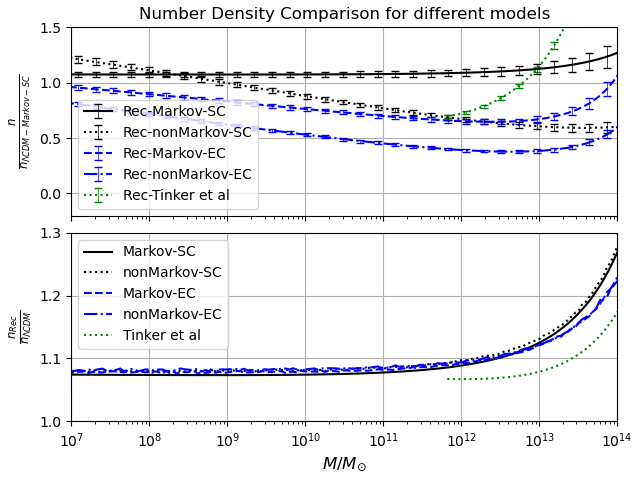}
	\caption{ Top panel: The ratio of DM halos' number density in the reconstructed model (Markov and non-Markov) for spherical and ellipsoidal collapse and the simulation based fitting function of Tinker et al. \citep{Tinker:2008ff} to Planck-2018 $\Lambda$CDM Markov model with spherical collapse case is plotted in redshift $z=0.5$. Bottom panel: The ratio of the reconstructed model to Planck-2018 $\Lambda$CDM model is plotted in redshift $z=0.5$ for each case of EST and collapse model, separately. }\label{fig5-comp}
\end{figure}
\begin{figure}
	\includegraphics[scale=0.5]{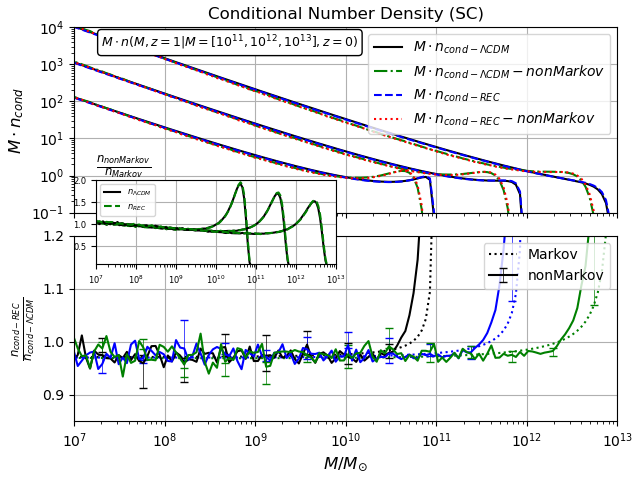}
	\caption{Top panel: The mass function of DM progenitors in redshift $z=1$, which will merge to form halos with masses $10^{11},10^{12},10^{13}M_{\odot}$ in present time $z=0$ is plotted from left to right. Bottom panel: The ratio of conditional mass function predicted by reconstructed model to Planck-2018 $\Lambda$CDM is plotted versus redshift for Markov to Markov (dashed line) and non-Markov to non-Markov (solid line) cases, separately. The three curves are related to the three final masses $10^{11},10^{12},10^{13}M_{\odot}$ from left to right. Inset figure: The ratio of the non-Markov to Markov is plotted. For figure clarity the error-bars of the Markov case are not shown in the bottom panel. Note that all results are for spherical collapse model.} \label{fig5}
\end{figure}
\begin{figure}
	\includegraphics[scale=0.5]{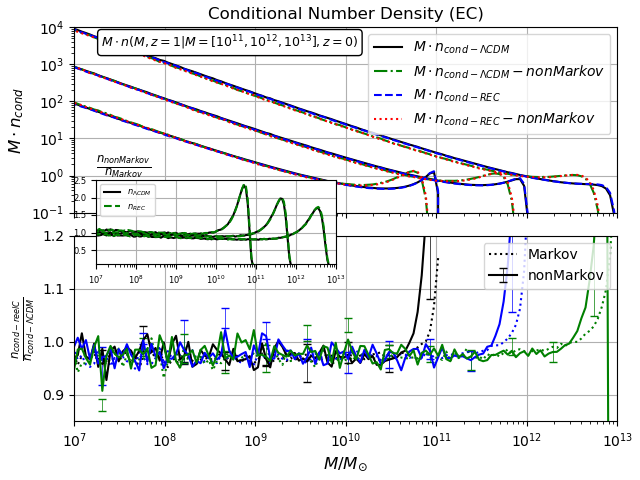}
	\caption{ Top panel: The mass function of DM progenitors in redshift $z=1$, which will merge to form halos with masses $10^{11},10^{12},10^{13}M_{\odot}$ in present time $z=0$ is plotted from left to right. Bottom panel: The ratio of conditional mass function predicted by reconstructed model to Planck-2018 $\Lambda$CDM is plotted versus redshift for Markov to Markov (dashed line) and non-Markov to non-Markov (solid line) cases, separately. The three curves are related to three final masses $10^{11},10^{12},10^{13}M_{\odot}$ from left to right. Inset figure: The ratio of the non-Markov to Markov is plotted. For figure clarity the error-bars of the Markov case are not shown in the bottom panel. Note that all results are for ellipsoidal collapse model.} \label{fig5ec}
\end{figure}
In top panel of Fig.\ref{fig5-comp}, we compare the number density of DM halos in $z=0.5$ in the reconstructed model (Markov and non-Markov) and (spherical and ellipsoidal collapse  Sheth et al. \cite{Sheth:2001dp,Sheth:2001})  and also the simulation-based fitting functions Tinker et al. model \cite{Tinker:2008ff}) to Planck-2018 $\Lambda$CDM Markov-spherical collapse model as the most basic analytically predicted DM number density. The top panel is shown to indicate the dependence of the number density of dark matter halos to each of the models discussed above. In the bottom panel of Fig.\ref{fig5-comp}, we compare the ratio of the reconstructed cosmological model to Planck-2018 $\Lambda$CDM for each model of EST (Markov/non-Markov) and collapse model (SC/EC) in $z=0.5$. This plot shows that the ratio of the two cosmological models is almost independent of the EST and collapse models. This is because of the cancellation of the changes in numerator and dominator of the ratios due to structure formation models. Accordingly, the main effect is due to the change of the Hubble parameter.

Also, the scope of the validity of our assumption, which relates the first up-crossing statistics to the DM halo number density straightforwardly, must be reconsidered. In this direction, ideas such as peak theory \cite{Bardeen:1985tr} and the excursion set theory of peaks \cite{Paranjape:2012ks} and more recently, excursion set peaks in energy have been introduced \cite{Musso:2019zmr}. \\
Furthermore, to compare our results with observational data, we should consider all the complications raised from the halo occupation distribution physics \cite{Mo2010}. That leads us to predict an observable change in galaxies' luminosity function compared to the standard model.
\begin{figure}
\includegraphics[scale=0.5]{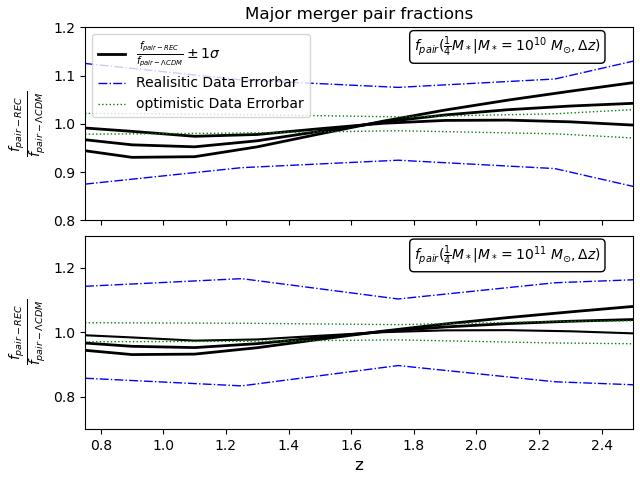}
\caption{ The conditional mass fraction of DM halo progenitor of the stellar mass $\frac{1}{4}M_*$ which forms a halo with stellar mass $M_*$ (dark matter halo mass $M=10M_*$) in the redshift interval $\Delta z$ is plotted versus redshift (solid black lines with $1\sigma$ interval). This plot shows the ratio of this quantity in the reconstructed model to the Planck-2018 $\Lambda$CDM, Markov case in SC. The ratio is plotted for the final stellar Mass $M_*=10^{10}M_{\odot}$ in the top panel and $M_*=10^{11}M_{\odot}$ in the bottom panel. The dash-dotted blue lines show the realistic error bars on the pair fraction derived from Table 2 of \citep{Duncan2019} and the green dotted lines show an optimistic future error bar with the 5 times improvement.} \label{fig7}
\end{figure}
To overcome this obstacle, we suggest that DM halos' mass assembly history can be related to an observational quantity such as pairing fraction of the galaxies \cite{Duncan2019}.
One important quantity, related to mass assembly history is the progenitor distribution of DM halos. In Fig.\ref{fig5}, we plot the mass function of DM progenitors in redshift $z=1$, which will merge to form halos with masses $10^{11},10^{12},10^{13}M_{\odot}$ in present time $z=0$ (see equation(\ref{eq:ffuc})) in the context of SC. The bottom panel of Fig.\ref{fig5} shows the ratio of conditional mass function predicted by the reconstructed model to the Planck-2018 $\Lambda$CDM in SC for Markov to Markov (dashed line) and non-Markov to non-Markov (solid line) cases, separately. The three curves are related to three final masses $10^{11},10^{12},10^{13}M_{\odot}$ from left to right. The inset figure shows the ratio of the non-Markov to Markov case for both cosmological models in SC.
In Fig.\ref{fig5ec}, the same results of Fig.\ref{fig5} is shown for EC.
{{The theoretical uncertainty of progenitor mass distribution predicted by the reconstructed model is too large to distinguish the models. However, the difference between two models become important  when the mass of the progenitor is near to the final mass, which corresponds to the major mergers.}}
We should note that the Markov and non-Markov behavior is almost the same for SC and EC.\\
In a recent work by \cite{Duncan2019}, a study is done to find the major merger pair fraction\footnote{Mass ratio of the major merger is in the range of > 1/4} of galaxies in the Hubble Space Telescope (HST) Cosmic Assembly Near-infrared Deep Extragalactic Legacy Survey (CANDELS). The idea is based on counting the pair galaxies in redshift bins up to $z \simeq 6.5$. The galaxies distance separation in the interval of $r \simeq (5-30) \text{kpc}$ and redshift or velocity separation of $\Delta V \leq 500 \text{ km s}^{-1}$ are considered as candidates for merging. The quantity $f_{\text{pair}} = (\text{Number of pair galaxies} ) / (\text{Number of total galaxies}) $ is defined respectively.
The major merger pair fraction of galaxies are obtained and reported for two stellar mass range $9.7 < \log M_* / M_{\odot} < 10.3$ and $ \log M_* / M_{\odot}>10.3 $ in Table 2 of \cite{Duncan2019}. If we assume a constant stellar mass to DM halo mass ratio (i.e $M_*/M = 0.1$) for the pairing galaxies, we can use this results to compare with DM halo major merger pair fraction, related to the conditional mass function.\\
We should emphasize that equation \eqref{eq:ffuc} calculate both the accretion and merger in the mass interval. Accordingly, we introduce a $\gamma$ parameter which shows the ratio of mass increment by merger to total mass assembly (merger and accretion). We assume that $\gamma$ parameter has a weak dependence to the redshift and cosmological model. Our result is the ratio of the pair galaxy's fraction in two models. So, hereafter we approximate the major merger pair fraction in EST framework by $f_{\text{merger}}=\gamma \times f_{FU}(S_1,\delta_1|S_2,\delta_2)$.
{{Note that the constant $\gamma$ is a rough simple assumption. The dependency of the $\gamma$ to redshift and cosmology can be an independent study which is not in the scope of this work. Accordingly, we should note that discussion on the relation of the mass progenitor distribution and observations is a mere proposal.  }}\\
In Fig.\ref{fig7}, the conditional mass fraction of DM halo progenitor  $\frac{1}{4}M_*$ which forms a halo with stellar mass $M_*$ (dark matter halo mass $M=10M_*$) in the redshift interval $\Delta z$ is plotted versus redshift (solid black lines with $1\sigma$ interval).
 This plot shows the ratio of this quantity in the reconstructed model to the Planck-2018 $\Lambda$CDM, Markov case in SC. Note that the results are almost similar for non-Markov and EC cases as we show in previous figures. The ratio is plotted for the final stellar Mass $M_*=10^{10}M_{\odot}$ in the top panel and $M_*=10^{11}M_{\odot}$ in the bottom panel. The dash-dotted blue lines show the realistic error bars on the pair fraction derived from Table 2 of \citep{Duncan2019} and the green dotted lines show the optimistic future error bars with the 5 times improvement. 
{{The error-bars corresponds to reconstructed Hubble constant errors and the related $D(z)$. 
The deviation of the ratio of two models from unity is larger  in low redshift $\lesssim 1.2$ and higher redshift $z\gtrsim 2.0$ which give the opportunity to distinguish two models in optimistic case. However, due to discussion on the complications of $\gamma$ factor and the assumption of constant $\gamma$, at the moment Fig.\ref{fig7}  is for illustration only, since the theoretical uncertainty of the result may be well larger than the observational error.}}\\
The optimistic case  can be reached by increasing the precision of the observations with better photometric and spectroscopic measurements. Also, the improvement can be achieved by increasing the statistics of galaxies by future LSS surveys. We assert that one can distinguish the standard $\Lambda$CDM from the reconstructed model by future observations.
We obtain all results independent of the dark energy model, and it can be easily applied to any other cosmological model, which affects the Hubble parameter.

\section{Conclusion and Future remarks}
\label{Sec:4}
The standard cosmological model known as $\Lambda$CDM is very successful in describing different observations from CMB to the galaxies' distribution in the late time. However, there are observational and theoretical tensions, which may introduce new venues to go beyond the standard model and shed light on the physics of dark energy, dark matter, and the early Universe. The $H_0$ tension is one of the most challenging and discussed problems in recent years. In this work, we use a cosmological background evolution model-independent reconstructed Hubble parameter based on late time background observational data (i.e., SNe Ia, BAO, CMB distance indicator, and Hubble constant from local measurements), as an alternative to $\Lambda$CDM. Based on this model, we calculate the LSS observables, such as the number density of DM halos, the probability distribution of DM progenitors, and the mass assembly history of DM halos. We compare the results with the Planck-2018 $\Lambda$CDM. This procedure's idea is that the LSS observables in non-linear scales can be used as a further criterion to distinguish the models, which could shed more light on the Hubble tension. We are interested in the mass assembly history of dark matter halos as the hierarchical structure formation's backbone. \\
{{We show that the error bars of the reconstructed model are as much as we can distinguish the two models by their DM halo number density prediction approximately in $\sim 2\sigma$ in DM halo mass ranges  $M\gtrsim10^{12}M_{\odot}$. The theoretical uncertainty of progenitor mass distribution predicted by the reconstructed model is too large to distinguish the models. However, the difference between two models become important for major mergers.}} It should be noted that these results are in the context of EST Markov and non-Markov extension, considering the SC and EC cases. These results motivate us to develop N-body simulations based on different Hubble parameter histories (e.g., introduced the reconstructed model) for more accurate results. However, to compare the suggested probes with observational data, we should consider all the complications raised from the physics of the halo occupation distribution. To find relations between the DM host halos' statistics and mass assembly history and the luminosity, color, and morphology distribution of the galaxies.
We are inspired by an exciting observation \cite{Duncan2019} with Hubble space telescope HST, CANDEL field on the number of ``very near galaxy pairs'' as an indication of merging galaxies. We calculate the ratio of the major mass assembly fraction of DM progenitors (for two cosmological models of Planck-2018 $\Lambda$CDM and reconstructed model) in the excursion set theory context in Markov and non-Markov extensions with considering both spherical and elliptical collapse models.
To be more specific, we propose that the conditional mass fraction of DM halos (as shown in Fig.\ref{fig7}) is related to the statistics of pair galaxies in the HST, CANDEL Survey. We show decreasing the error bars on the pair statistics can distinguish the models with future observations ($\sim 5$ times more accurate data). Better precision can be achieved with better photometric, spectroscopic measurements, and increasing the statistics of galaxies.\\
For the future, this work can be restudied in the context of peak theories  \cite{Cadiou:2020xmo} and N-body simulations. The physics of halo occupation distribution and luminous matter's bias to dark matter should be reconsidered in the alternative models. 
{{Also the cosmology and redshift dependence of the $\gamma$ parameter which we introduce as the fraction of mass increment in major merger to total mass assembly (merger and accretion), could be the subject of further studies. Due to discussion on the complications of $\gamma$ factor and the assumption of constant $\gamma$, at the moment our result on pair fraction is for illustration only, since the theoretical uncertainty of the result may be well larger than the observational error.}} \\ We should note that we study the models that differ from $\Lambda$CDM only in the Hubble parameter in this scheme. There are theories (e.g., modified gravity) that change the physics of the collapse and the Poisson equation. For this category of models, the non-linear structure formation (collapse models and EST, ...) should be reformulated.
Finally, we emphasize that by the upcoming LSS surveys such as Euclid, Vera C. Rubin Observatory, Nancy Grace Roman Space Telescope, ... we will have the opportunity to test models in both background evolution and linear (non-linear) structure formation.
\section*{Acknowledgments}
We are grateful to Levon Pogosian, providing us with the reconstructed Hubble parameter data.
We thank Ravi K. Sheth and Nima Khosravi for valuable discussions.
We thank the anonymous referee for his/her insightful comments and suggestion, which improve the manuscript extensively.
SB is partially supported by Abdus Salam International Center of Theoretical Physics (ICTP) under the junior associateship scheme. This research is supported by Sharif University of Technology Office of Vice President for Research under Grant No. G960202. \\ \\
\section*{Data Availability}
The results are reported based on the data provided in\cite{Wang:2018fng} for the reconstructed Hubble parameter and \cite{Duncan2019} for the galaxy major merger. The data for all plots based on our theoretical models are available upon request.



\bsp	
\label{lastpage}
\end{document}